\documentclass[pra,twocolumn,nofootinbib,floatfix]{revtex4-1} 
\usepackage[usenames]{color}
\usepackage{hyperref}

\usepackage{bbold}

\usepackage{hhline}

\usepackage[utf8]{inputenc}
\usepackage[T1]{fontenc}
\usepackage[pdftex]{graphicx}

\usepackage{amsmath, amsxtra, amssymb, mathtools, isomath, xspace, textcomp} 

\usepackage[british]{babel}

\usepackage{booktabs}
\usepackage{multirow}

\newcommand{\mycite}[1]{\cite{#1}}

\newcommand{\ket}[1]{\ensuremath{\vert #1 \rangle}\xspace}%
\newcommand{\bra}[1]{\ensuremath{\langle #1 \vert}\xspace}%

\newcommand{\down}{\mbox{`$\,\downarrow\,$'}}
\newcommand{\up}{\mbox{`$\,\uparrow\,$'}}

\long\def\symbolfootnote[#1]#2{\begingroup%
\def\thefootnote{\fnsymbol{footnote}}\footnotetext[#1]{#2}\endgroup}

\begin{document}

\title{Quantum dynamics of a single, mobile spin impurity}


\author{Takeshi Fukuhara$^{1,*}$}%
\author{Adrian Kantian$^{2}$}%
\author{Manuel Endres$^{1}$}%
\author{Marc Cheneau$^{1}$}%
\author{Peter Schau{\ss}$^{1}$}%
\author{Sebastian Hild$^{1}$}%
\author{David~Bellem$^{1}$}%
\author{Ulrich Schollw\"{o}ck$^{3}$}%
\author{Thierry Giamarchi$^{2}$}%
\author{Christian Gross$^{1}$}%
\author{Immanuel Bloch$^{1,3}$}%
\author{Stefan Kuhr$^{1,4}$}%


\affiliation{$^1$Max-Planck-Institut f\"{u}r Quantenoptik, Hans-Kopfermann-Str. 1, 85748 Garching, Germany}
\affiliation{$^2$DPMC-MaNEP, University of Geneva, 24 Quai Ernest-Ansermet, 1211 Geneva, Switzerland}%
\affiliation{$^3$Fakult\"{a}t f\"{u}r Physik, Ludwig-Maximilians-Universit\"{a}t M\"{u}nchen, 80799 M\"{u}nchen, Germany}%
\affiliation{$^4$University of Strathclyde, Department of Physics, SUPA, Glasgow G4 0NG, United Kingdom}%

\begin{abstract}
Quantum magnetism describes the properties of many materials such as transition metal oxides and cuprate superconductors. One of its elementary processes is the propagation of spin excitations. Here we study the quantum dynamics of a deterministically created spin-impurity atom, as it propagates in a one-dimensional lattice system. We probe the full spatial probability distribution of the impurity at different times using single-site-resolved imaging of bosonic atoms in an optical lattice. In the Mott-insulating regime, a post-selection of the data allows to reduce the effect of temperature, giving access to a space- and time-resolved
 measurement of the quantum-coherent propagation of a magnetic excitation in the Heisenberg model.
Extending the study to the bath's superfluid regime, we determine quantitatively how the bath strongly affects the  motion of the impurity. The experimental data shows a remarkable agreement with theoretical predictions allowing us to determine
the effect of temperature on the coherence and velocity of impurity motion. Our results pave the way for a new approach to study quantum magnetism, mobile impurities in quantum fluids, and polarons in lattice systems.
\end{abstract}

  \symbolfootnote[1]{Electronic address: {\bf takeshi.fukuhara@mpq.mpg.de}}
  \maketitle%


Deepening our knowledge of quantum magnetism \mycite{Auerbach:1998} is one of the most important goals in quantum simulation.
In particular, one highly desired aim is a better understanding of high-T$_c$ cuprate superconductors, which are believed to be described by Heisenberg-type effective spin models in the limit of low doping \mycite{Book:LesHouches1995}. More generally, the physics characterized by the Heisenberg model governs the properties of many strongly correlated materials
such as transition metal oxides \mycite{Imada:1998}, and allows the realization of various remarkable types of spin order ranging from spin solids to spin liquids
\mycite{Auerbach:1998,Balents:2010}. In low-dimensional quantum magnets, particularly rich physics emerges due to the dynamics of spin excitations \mycite{Cazalilla:2011}. A basic mechanism of quantum magnetism in strongly correlated electronic systems is superexchange, in which opposite spins on adjacent lattice sites coherently exchange their positions. Ultracold atoms in optical lattices offer an ideal testbed to explore these phenomena in a controlled experimental environment \mycite{Bloch:2008c}, as shown by the observation of
superexchange in double-well systems \mycite{Trotzky:2008a} or plaquettes~\mycite{Nascimbene:2012}. Furthermore, the recently demonstrated single-atom-resolved detection \mycite{Bakr:2010,Sherson:2010} opens entirely new prospects for the quantum simulation of strongly correlated spin systems. For example, this technique has enabled the simulation of one-dimen\-sional anti-ferromagnetic Ising spin chains via mapping of the site occupation onto a pseudo-spin \mycite{Simon:2011}. In ultracold-atom experiments, however, observation of superexchange-based quantum magnetism in many-body systems
has so far been hindered by the fact that typical temperatures are much larger than the superexchange coupling energy.

\begin{figure}[!t]
    \centering
     \includegraphics[width=\columnwidth]{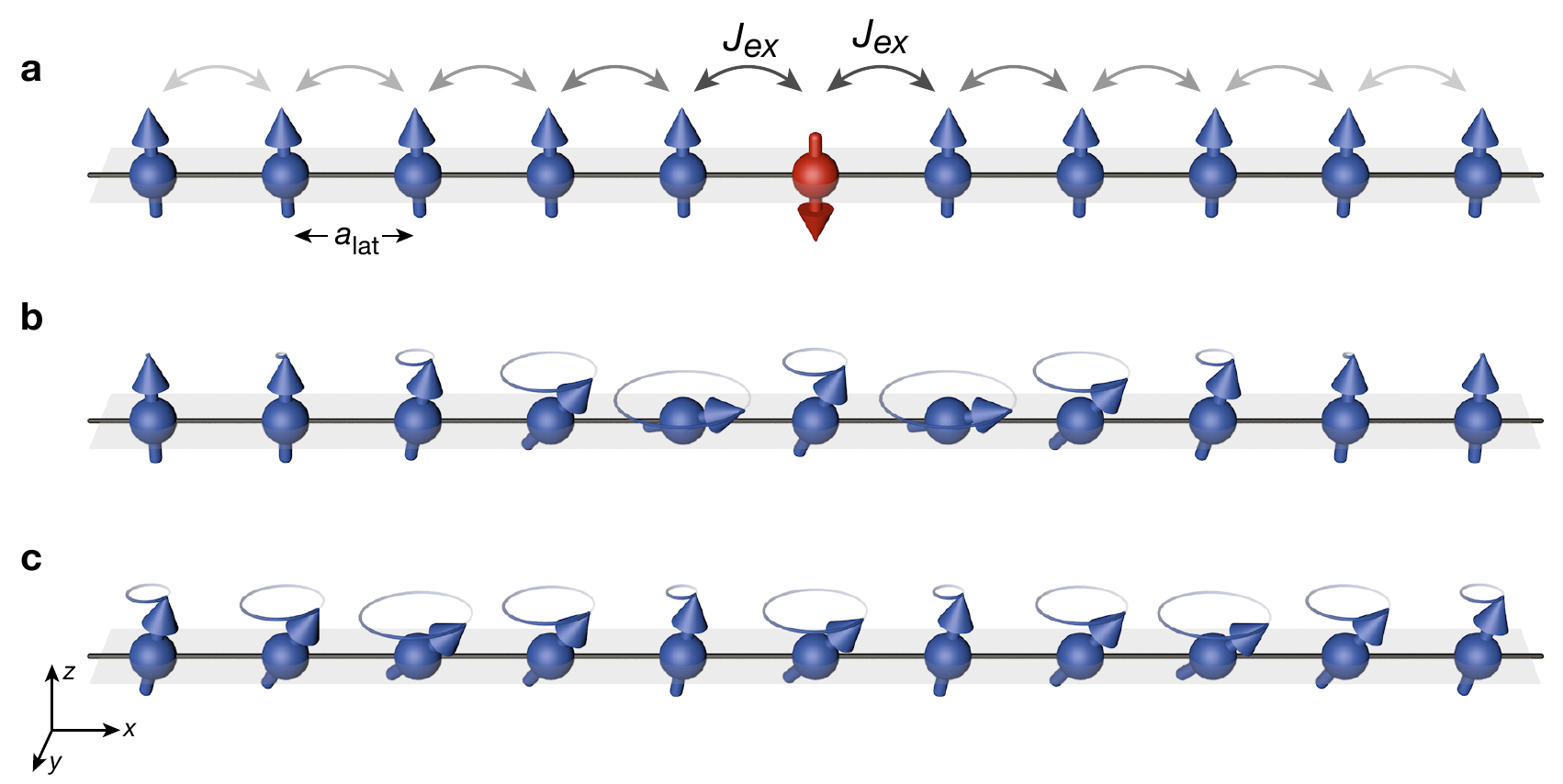}
    \caption{{\bf Coherent propagation of a single spin excitation in the Heisenberg model.} {\bf a}, A single spin is flipped at the centre site of a 1D spin chain. Each spin is coherently coupled to its neighbors via the superexchange coupling $J_{\rm ex}$. {\bf b}, {\bf c} Coherent evolution showing the polar angles of the spins after times $2 \hbar / J_{\rm ex}$ and $4 \hbar/J_{\rm ex}$. Circles visualize the quantum uncertainty of the transverse spin component.}
    \label{fig:schematics}
\end{figure}

In this work, we report on the first space- and time-resolved observation of a coherently propagating spin-wave, by tracking the motion of a deterministically created single-spin impurity in a one-dimensional (1D) spin chain (Fig.\,\ref{fig:schematics}).
 Specifically, we studied the quantum dynamics of a mobile boson of type \down\ on a 1D lattice, surrounded by a bath of bosons of type \up. Such a system can be described within a two-species single-band Bose-Hubbard model, parametrized by the spin-independent single-particle tunneling rate $J$ and on-site interaction energy $U$ (Supplementary Information). Deep in the Mott insulator (MI) regime ($U \gg J$) with unity filling, this model can be mapped to the isotropic Heisenberg model \mycite{Kuklov:2003,Duan:2003,GarciaRipoll:2003,Altman:2003a}:
\begin{subequations}
\begin{align}
\hat H  &=  - J_{\rm ex} \sum_{\langle j,k \rangle} \mbox{\boldmath $\hat S$}_j \cdot \mbox{\boldmath $\hat S$}_k\label{eq:Heisenberg0}\\
  &=  - \frac{J_{\rm ex}}{2} \sum_{\langle j,k \rangle} \left( \hat S_j^+ \hat S_k^- + \hat S_j^- \hat S_k^+ \right) - J_{\rm ex} \sum_{\langle j,k \rangle} \hat S_j^z \hat S_k^z ,
\label{eq:Heisenberg1}
\end{align}
\end{subequations}
in which the effective superexchange coupling $J_{\rm ex}  = 4 J^2/U$ arises from a second-order tunneling process. The bosons \up\ and \down\ are identified with spin states $\ket{\uparrow}$ and $\ket{\downarrow}$ and the corresponding operators of the effective spin system are $\mbox{\boldmath $\hat S$}_j=(\mbox{$\hat S^x_j$},\mbox{$\hat S^y_j$},\mbox{$\hat S^z_j$})$,  $\hat S_j^+ = \hat S_j^x + i \hat S_j^y = \ket{\uparrow}_j \bra{\downarrow}_j$\,, $\hat S_j^- = \hat S_j^x - i \hat S_j^y = \ket{\downarrow}_j \bra{\uparrow}_j$ and $\hat S_j^z = \left(\hat{n}_ {\uparrow ,j} - \hat{n}_ {\downarrow ,j} \right) / 2$, with the number operators $\hat{n}_ {\sigma ,j}$  for bosons of type $\sigma =\ \uparrow, \downarrow$ on lattice site $j$. In the case of a single flipped spin in a ferromagnetic domain, the second term of Eq.\,(\ref{eq:Heisenberg1}), which describes the longitudinal spin coupling, only gives rise to an energy offset and can therefore be neglected. The remaining first term of Eq.\,(\ref{eq:Heisenberg1}) is structurally equivalent to the single-species single-particle tunneling Hamiltonian within the tight binding model of a 1D lattice system \mycite{Book:FeynmanStatMech1972,Auerbach:1998,Cazalilla:2011}, with the tunneling rate $J$ being replaced by $J_{\rm ex}/2$, and the atomic creation and annihilation operators being replaced by the spin-flip operators $\hat S_j^+$ and $\hat S_j^-$.

For larger $J/U$, the mapping to the Heisenberg model breaks down due to the fluctuations of the site occupancies and due to higher-order processes, and no analytic solution describing the time evolution of the impurity exists. In this regime, the motion of the impurity is modified compared to both a free particle and an isolated magnetic excitation. This effect can be interpreted as a result of interactions between the impurity and low-lying density excitations, which resemble phonons, leading to polaron-like effects.


In order to create 1D spin chains in the optical lattice, we followed the experimental procedure of our previous work \mycite{Endres:2011}. We started by preparing a two-dimensional degenerate gas of about 170 $^{87}$Rb atoms in a single antinode of a vertical optical lattice (lattice spacing $a_{\rm lat}=532$\,nm) along the  $z$ direction. We then switched on two horizontal optical lattice beams within $120$\,ms with potential depths $V_x=10.0(3)\,E_r$ and $V_y=30(2)\,E_r$ (the number in parenthesis denotes the uncertainty of the last digit), where $E_{r}=h^2/(8m a_{\rm lat}^2)$ denotes the recoil energy and $m$ is the atomic mass of $^{87}$Rb. This creates an array of parallel 1D MIs along the $x$ direction, containing each 8 to 16 atoms. The atoms were initially prepared in the hyperfine state $\ket{F=1, m_F=-1}$ ($\equiv$ \up). We introduced the spin impurity by changing the hyperfine state of an  atom at the centre of the chain to state $\ket{F=2, m_F=-2}$ ($\equiv$ \down) using single-site addressing \mycite{Weitenberg:2011}. In this scheme, a $\sigma^-$-polarized, off-resonant laser beam at $787.65$\,nm  wavelength focussed onto the selected lattice site results in a negative energy shift (attractive potential) only for the state \down\ while leaving the initial spin state \up\ almost unaffected. A microwave pulse, resonant with the shifted atomic transition, then produced the spin-flip from \up\ to \down. In contrast to our previous work \mycite{Weitenberg:2011}, a spatial light modulator generated an addressing beam profile in form of a line instead of a circular-shaped Gaussian beam, in order to create the impurity in all 1D chains simultaneously (Methods). This novel multiple-site addressing technique offers the advantage to prepare an arbitrary spin pattern in the lattice more rapidly.
In addition, the line-shaped beam allowed us to hold the impurities in all 1D systems, and to release them simultaneously by switching off the beam within 1 ms. We then allowed the impurity in each chain to propagate during a  variable hold time, and finally froze the dynamics by rapidly increasing the lattice depth of all axes to $> 80\,E_r$ within $300$\,\textmu s.

The resulting spin distribution was detected by single-site-resolved fluorescence imaging using a high-resolution microscope objective \mycite{Sherson:2010}. We used two alternative detection methods to determine the position of the impurity: (i) the direct imaging of the impurity spin component \down\ after removing all atoms in state \up\ with a resonant laser pulse  (`positive image') and (ii) detecting it as an empty site in the bath of \up -atoms after removing the spin impurity (`negative image'). The latter has the advantage to provide information about the thermal excitations (holons and doublons) of the system as well, which are also detected as empty sites. Although we cannot distinguish between spin impurity and thermal excitations, post-selecting  samples with only one empty site in the chain enables us to filter out a lower-temperature subset of the data. If the only empty site arises from a spin-flipped atom, the position of the impurity can be determined exactly and the selected samples contain no additional excitations.

\begin{figure}
    \centering
     \includegraphics[width=\columnwidth]{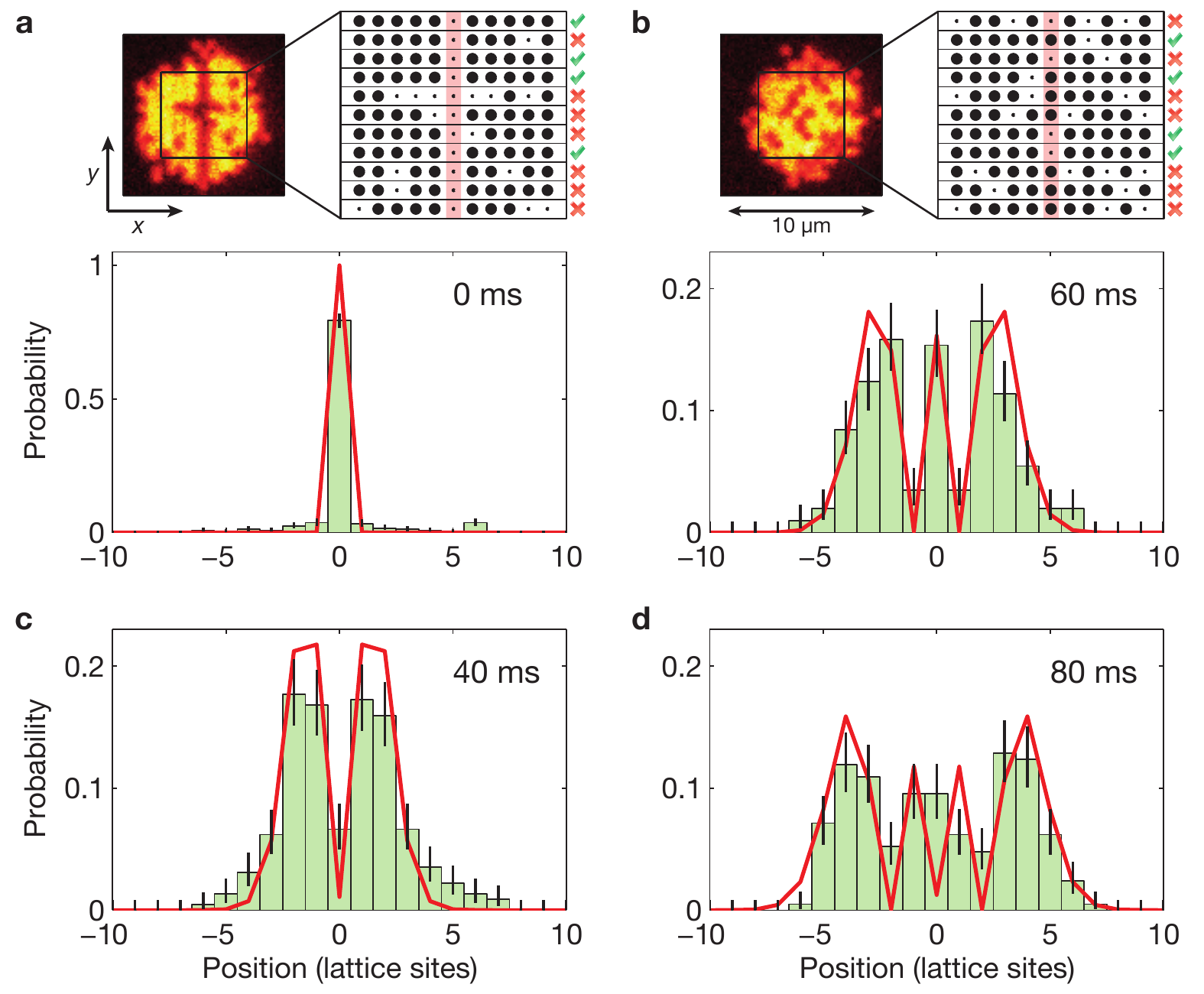}
    \caption{{\bf Dynamics of a mobile spin impurity.} The top panels in {\bf a} and {\bf b} show fluorescence images of the atoms (left) taken with the `negative' imaging technique (see main text) together with the reconstructed atom distribution in the central region (right). The 1D systems are oriented horizontally. The red vertical stripe denotes the initial position of the spin impurity (detected as an empty lattice site). For the generation of an effective low-temperature subset, only the samples containing exactly one empty site were kept (green tick marks), while those containing more than one empty site were discarded (red crosses). The histograms in {\bf a}-{\bf d} show the position distribution of the spin impurity  after different hold times for $J/U=0.053$.  Each histogram is obtained from an average over 200-250 1D systems. The error bars denote the $1\sigma$ statistical uncertainty. The red line is a simultaneous fit to all distributions with the analytic solution of the Heisenberg model of equation (\ref{eq:ProbabilityDistribution}), yielding $J_{\rm ex}/\hbar= 65(1)$\,Hz.}
    \label{fig:coherent_dynamics}
  \end{figure}
\begin{figure*}[!t]
    \centering
     \includegraphics[width=0.95\textwidth]{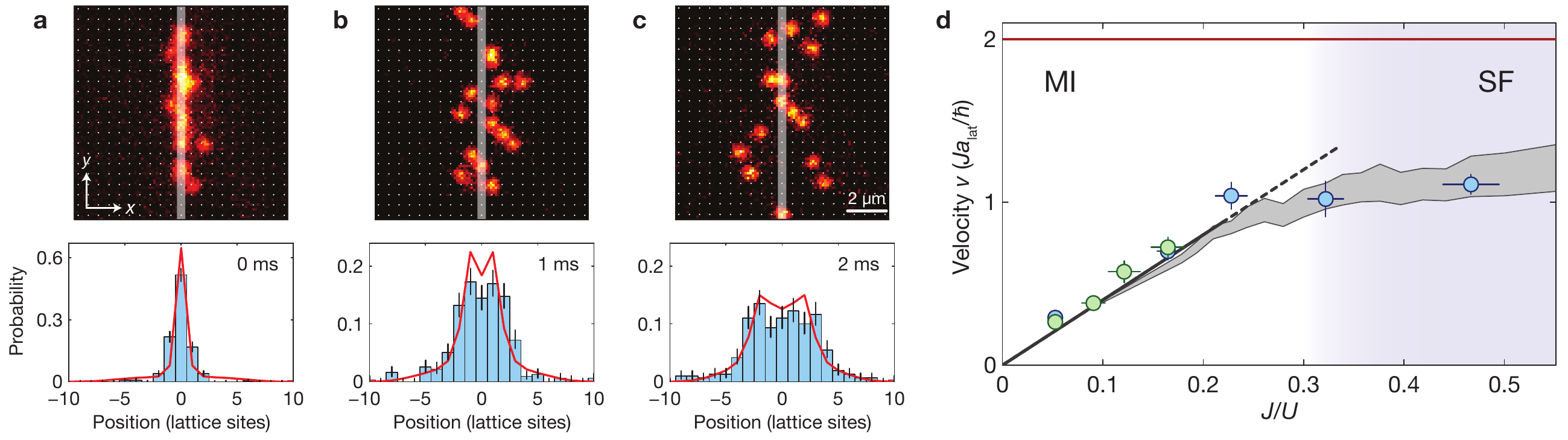}
    \caption{{\bf Spin dynamics across the superfluid-to-Mott-insulator transition.}
{\bf a}-{\bf c}, Spin impurity dynamics in the SF regime $\left(J/U = 0.32\right)$ close to the critical point $\left[(J/U)_c \approx 0.3\right]$ for different hold times. The upper panels display fluorescence images of the impurity spins after removing the other spin component (`positive image'). The 1D chains containing more than one atom were excluded from the data analysis. The white vertical stripe highlights the initial position of the flipped spin. The lower panels show the position distribution averaged over about 300 chains (blue bars) together with a t-DMRG simulation at $T=0.11\,U/k_{\rm B}$ (red line).
{\bf d}, Measured velocities of the spin impurity for different values of $J/U$ extracted from negative (green circles) and positive images (blue circles); horizontal and vertical error bars indicate the 1$\sigma$ uncertainties of the lattice depth and the combination of fit error and uncertainties of $J$, respectively.
The dark gray line shows scaling with $4 J^2/U$, whereas the brown line indicates the propagation velocity of a single free particle $\left(J/U=\infty\right)$. The gray shaded region shows results from a t-DMRG simulation at $T=0$ taking into account varying initial atom numbers. The area denotes the 1$\sigma$ fit error to the simulated distributions.}
    \label{fig:spreading_speed}
  \end{figure*}

We first studied the time evolution of a spin impurity deep in the MI regime using negative images. The probability distribution of its position was obtained by averaging data from different chains, post-selected to contain only one empty site within the central 10-14 sites. The distributions [see Fig.\,\ref{fig:coherent_dynamics} for $J/U=0.053(7)$] show clear maxima and minima, resulting from the quantum interference due to the coherent evolution of the spin impurity. Because of the weaker superexchange coupling, the observed dynamics occurs on timescales much longer than the tunneling time ($\hbar/J = 4$\,ms) that would characterize the motion of non-interacting atoms. We compared our data to the time evolution of the spin impurity in the exactly solvable homogeneous Heisenberg model at zero temperature. In this model, the probability of finding the impurity at time $t$ on site $j$ after starting its evolution at $t=0$ from the centre of the chain ($j=0$) is
\begin{equation}
P_j \left( t \right) = \left[ {\cal J}_j \left( \frac{J_{\rm ex} t}{\hbar} \right) \right]^2, \label{eq:ProbabilityDistribution}
\end{equation}
where ${\cal J}_j$ is the Bessel function of the first kind \mycite{Konno:2005}. A single fit to all distributions observed at different hold times (red curves in Fig.\,\ref{fig:coherent_dynamics}) with the superexchange coupling as a free parameter yields good agreement with the data for $J_{\rm ex}/\hbar =65(1)$\,Hz, which is close to $J_{\rm ex}/\hbar=51(^{+11}_{-8})$\,Hz obtained from an ab-initio band-structure calculation using the independently measured lattice depths (see also Fig.\,\ref{fig:spreading_speed}). Small differences between the data and the model result from the limited efficiency of the post-selection process (Supplementary Information). We note that the effect of the external potential on the spin dynamics can be neglected in the MI limit, as long as the impurity has not yet reached the edge of the system (Supplementary Information).


  \begin{figure}[!b]
    \centering
     \includegraphics[width=\columnwidth]{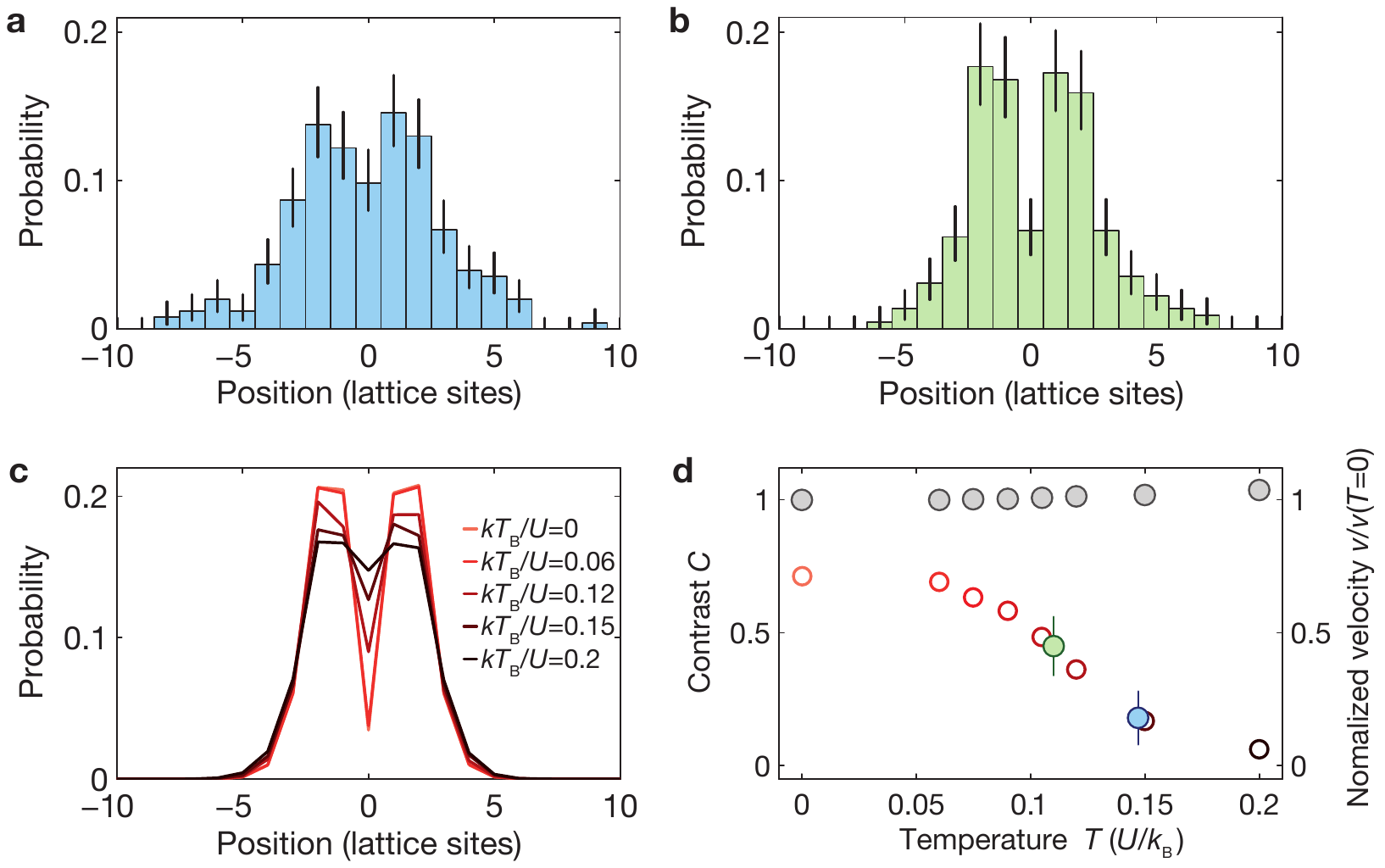}
    \caption{{\bf Effect of thermal excitations on the coherent spin dynamics.} Shown are position distributions of the spin impurity for $J/U = 0.053$ and a hold time of $40$\,ms, {\bf a} generated from positive images (blue bars) and {\bf b} after post-selection from negative images containing one empty site in the chain (green bars). {\bf c}, Position distributions from t-DMRG simulations for different temperatures. {\bf d}, Contrast of the simulated distributions (open circles) and propagation velocity (gray filled circles) as a function of the temperature $T$. The blue and green filled circles show the contrast extracted from the experimental data of {\bf a} and {\bf b}, respectively. Error bars denote the 1$\sigma$ uncertainty.}
    \label{fig:excitation_effect}
  \end{figure}


To examine the effects of temperature, we measured the same distribution using positive imaging (Fig.\,\ref{fig:excitation_effect}a), in which we directly observed the impurity position in the finite-temperature bath. Compared with the distribution from the post-selected negative images with one empty site (Fig.\,\ref{fig:excitation_effect}b), it has almost the same width, but less contrast. This indicates that the propagation velocity of the spin impurity is not much affected by temperature, although its coherent evolution is hindered. Our experimental observation is quantitatively confirmed by  numerical simulations (Fig.\,\ref{fig:excitation_effect}c and d) using the time-dependent density-matrix renormalization group (t-DMRG) algorithm, including temperature (Methods). Our intuitive interpretation of this effect is that thermal excitations  can move back and forth  across the spin impurity, introducing phase slips that wash out the coherence of the distribution. However, a single excitation alters the position of the spin impurity at most by one site in a 1D system and as a consequence the width of the distribution is less affected.
For a quantitative analysis, we evaluated the contrast and the propagation speed from the simulated distributions for different temperatures from $T=0$ to $T=0.2\,U/k_{\rm B}$ (Fig.\,\ref{fig:excitation_effect}d). We defined the contrast as $C= (P_{\rm max}-P_{\rm min})/(P_{\rm max}+P_{\rm min})$,  where $P_{\rm max}$ is the peak value at positions other than the centre, and $P_{\rm min}$ denotes the minimum in the centre region between the peaks (i.e., $C=0$ if there is no other maximum except for the central peak).  The contrast obtained from the positive image  equals that of the simulated distributions at $T=0.15\,U/k_{\rm B}$ (blue point in Fig.\,\ref{fig:excitation_effect}d). This is consistent with the system's temperature of $T = 0.14(3)\,U/k_{\rm B}$, determined independently by comparing the distribution of thermal excitations to an analytical model in the atomic limit \mycite{Sherson:2010}. The contrast from negative images corresponds to $T = 0.11\,U/k_{\rm B}$ (green point in Fig.\,\ref{fig:excitation_effect}d), demonstrating that the post-selection enables us to extract a lower-temperature subset of the data. The actual temperature of this subset is, in fact, even lower, because our simulation does not take into account the finite spin-flip fidelity of 88(5)\%  and the fact that the initial position of the spin impurity could be empty due to a thermal excitation (see Supplementary Information).
We also determined the propagation speed of the spin impurity as a function of temperature by fitting the simulated distribution with the function of Eq.\,(\ref{eq:ProbabilityDistribution}), rewritten as $P_j \left( t \right) = \left[ {\cal J}_j \left( v t / a_{\rm lat} \right) \right]^2$, where we introduced $v$ as the speed of the impurity. We found that the change in velocity within the simulated temperature range is only 4\% (Fig.\,\ref{fig:excitation_effect}d). This result is consistent with our observation of similar propagation speeds from positive and negative images (see also Fig.\,\ref{fig:spreading_speed}d) and the intuitive argument mentioned above.



In the MI regime, the impurity propagates in an environment that is defect-free except for thermal excitations, leading to the coherent quantum motion described in the previous section.
The situation is radically different in the superfluid (SF) regime, in which the bath of \up\ spins contains many low-energy excitations as a result of quantum fluctuations of the on-site density.
In one dimension, those excitations are known to be phonon-like, which lead to a Fröhlich-type Hamiltonian now describing the drastic modification of the impurity motion in the bath by the propagation of a polaron \mycite{Book:Giamarchi}.
In order to investigate this regime, we measured the local impurity density at different times for increasing values of $J/U$. As the impurity is always prepared inside a Mott insulating state at $J/U = 0.053$, we decreased the lattice depth $V_x$ along the chain within $50$\,ms to reach the desired final value of $J/U$. While preparing the bath in this way, we kept the impurity spin pinned by the addressing beam before releasing it. For large final $J/U$, close to the SF-MI critical point, post-selecting samples with only one empty site is no longer effective for tracking the spin because of the proliferation of particle-hole pairs induced by quantum fluctuations \mycite{Endres:2011}. We therefore recorded the dynamics at large $J/U$ using positive images only, in which we directly detected the impurity spins (Fig.\,\ref{fig:spreading_speed}a-c).

Since there is no full analytical solution describing the dynamics of the spin impurity for intermediate values of $J/U$, we chose to analyze the experimental data by fitting the width of the position distribution with the function  $P_j \left( t \right)$ with the velocity
as an adjustable parameter. This function is indeed the analytic solution both for the Heisenberg (MI) regime ($J/U \ll 1$) with $v = J_{\rm ex} a_{\rm lat} / \hbar = 4 (J/U) \cdot (J a_{\rm lat} / \hbar)$ and for a totally free impurity ($J/U = \infty$) with $v =2 J a_{\rm lat} / \hbar $.
We found that a fit with $P_j \left( t \right)$ captures well the edges of the position distribution, allowing us to determine the maximum propagation velocity for all $J/U$ values used in this study.

We used the same fit to determine the propagation velocity from the numerical simulations. These took into account the statistical atom number fluctuations from the experiment as well as the trapping potential, which are both important in the SF regime. Over the whole range of $J/U$ accessed in the experiments, the measurement  shows a good agreement with the simulation, both for the position distribution (Fig.\,\ref{fig:spreading_speed}a-c), as well as for the impurity velocity (Fig.\,\ref{fig:spreading_speed}d). For small $J/U$, the experimentally determined spreading velocity normalized to the tunneling rate $J$ increases as $v/J = 4J/U$, as expected from the superexchange coupling $J_{\rm ex}  = 4 J^2/U$.

The spreading velocity becomes smaller than the velocity $v/J = 4J/U$ close to the point where the SF-MI transition would occur in a homogeneous system in the thermodynamic limit.
We interpret this change of velocity as a result from the interaction of the impurity with the increasingly large number of low energy excitations of the bath, which now affect not only the contrast but the velocity as well.
In the SF regime, experiment and numerical simulations show that the normalized velocity is only $v/J \approx a_{\rm lat}/\hbar$, i.e., it is only about half the velocity expected for free-particle tunneling. This indicates that the system's strong interactions in the superfluid regime still slow down the motion of the impurity significantly.  Such a behaviour is consistent with analytical calculations \mycite{Zvonarev:2007} and can be interpreted as a mass increase of the impurity which is expected from polaronic physics \mycite{Book:FeynmanStatMech1972,Schirotzek:2009,Nascimbene:2009,Catani:2012,Koschorreck:2012}.



In conclusion, we have observed the space- and time-resolved dynamics of a deterministically created spin impurity in a 1D lattice across the SF-MI transition.
By post-selecting measurements we filtered out a lower-temperature subset of our data, allowing for the observation of a coherently propagating magnetic excitation in the MI regime, consistent with magnetic superexchange.
Across the MI-SF transition our measurement  shows that the increasingly large number of low energy excitations of the bath affects both coherence and velocity of the impurity motion.
The velocity, which can be related to the renormalized tunneling amplitude in the lattice, is significantly reduced compared to a free particle. This is consistent
with the expected effect of such a bath and the mass increase due to polaronic effects, which is a field of intense ongoing research \mycite{Schirotzek:2009,Palzer:2009,Nascimbene:2009,Johnson:2011,Catani:2012,Koschorreck:2012,Schecter:2012,Spethmann:2012}.
The good agreement with numerical results demonstrates that our experiment is suitable for the quantitative study of mobile impurities in Bose-Hubbard models \mycite{Zvonarev:2009}. In particular, a predicted and fundamentally new universality class of single-particle excitations in 1D systems could be tested by measurement of spin-flip response and Green's function \mycite{Zvonarev:2007}. Other natural extensions of this work would be the investigation  of the correlated propagation of several flipped spins due to the ferromagnetic spin-spin coupling in the Heisenberg model \mycite{Ganahl:2012} or the quantum simulation of the impurity dynamics in 2D, where numerical simulations and analytical studies are much more difficult. Our multiple-site addressing technique could be used to prepare non-equilibrium states of quantum many-body systems such as domain walls \mycite{Gobert:2005} and study their evolution.
Further experiments could probe the fractionalization of excitations that naturally occurs in one dimension, either when magnons decay into two spinons, or when spin and charge excitations separate \mycite{Book:Giamarchi}.

    \section*{Acknowledgements}
We thank C. Weitenberg for his contribution to the single-site addressing scheme. This work was supported by MPG, DFG, EU (NAMEQUAM, AQUTE, Marie Curie Fellowship to M.C.), and JSPS (Postdoctoral Fellowship for Research Abroad to T.F.) and in part by the Swiss NSF under MaNEP and Division II.


  \bibliography{References}

\begin{thebibliography}{38}%
\makeatletter
\providecommand \@ifxundefined [1]{%
 \@ifx{#1\undefined}
}%
\providecommand \@ifnum [1]{%
 \ifnum #1\expandafter \@firstoftwo
 \else \expandafter \@secondoftwo
 \fi
}%
\providecommand \@ifx [1]{%
 \ifx #1\expandafter \@firstoftwo
 \else \expandafter \@secondoftwo
 \fi
}%
\providecommand \natexlab [1]{#1}%
\providecommand \enquote  [1]{``#1''}%
\providecommand \bibnamefont  [1]{#1}%
\providecommand \bibfnamefont [1]{#1}%
\providecommand \citenamefont [1]{#1}%
\providecommand \href@noop [0]{\@secondoftwo}%
\providecommand \href [0]{\begingroup \@sanitize@url \@href}%
\providecommand \@href[1]{\@@startlink{#1}\@@href}%
\providecommand \@@href[1]{\endgroup#1\@@endlink}%
\providecommand \@sanitize@url [0]{\catcode `\\12\catcode `\$12\catcode
  `\&12\catcode `\#12\catcode `\^12\catcode `\_12\catcode `\%12\relax}%
\providecommand \@@startlink[1]{}%
\providecommand \@@endlink[0]{}%
\providecommand \url  [0]{\begingroup\@sanitize@url \@url }%
\providecommand \@url [1]{\endgroup\@href {#1}{\urlprefix }}%
\providecommand \urlprefix  [0]{URL }%
\providecommand \Eprint [0]{\href }%
\providecommand \doibase [0]{http://dx.doi.org/}%
\providecommand \selectlanguage [0]{\@gobble}%
\providecommand \bibinfo  [0]{\@secondoftwo}%
\providecommand \bibfield  [0]{\@secondoftwo}%
\providecommand \translation [1]{[#1]}%
\providecommand \BibitemOpen [0]{}%
\providecommand \bibitemStop [0]{}%
\providecommand \bibitemNoStop [0]{.\EOS\space}%
\providecommand \EOS [0]{\spacefactor3000\relax}%
\providecommand \BibitemShut  [1]{\csname bibitem#1\endcsname}%
\let\auto@bib@innerbib\@empty
\bibitem [{\citenamefont {Auerbach}(1998)}]{Auerbach:1998}%
  \BibitemOpen
  \bibfield  {author} {\bibinfo {author} {\bibfnamefont {A.}~\bibnamefont
  {Auerbach}},\ }\href@noop {} {\emph {\bibinfo {title} {{Interacting Electrons
  and Quantum Magnetism}}}}\ (\bibinfo  {publisher} {Springer},\ \bibinfo
  {address} {Berlin},\ \bibinfo {year} {1998})\BibitemShut {NoStop}%
\bibitem [{\citenamefont {Dou\c{c}ot}\ and\ \citenamefont
  {Zinn-Justin}(1995)}]{Book:LesHouches1995}%
  \BibitemOpen
  \bibinfo {editor} {\bibfnamefont {B.}~\bibnamefont {Dou\c{c}ot}}\ and\
  \bibinfo {editor} {\bibfnamefont {J.}~\bibnamefont {Zinn-Justin}},\ eds.,\
  \href@noop {} {\emph {\bibinfo {title} {Proceedings of the Les Houches Summer
  School, Session LVI}}}\ (\bibinfo  {publisher} {Elsevier},\ \bibinfo
  {address} {Amsterdam},\ \bibinfo {year} {1995})\BibitemShut {NoStop}%
\bibitem [{\citenamefont {Imada}\ \emph {et~al.}(1998)\citenamefont {Imada},
  \citenamefont {Fujimori},\ and\ \citenamefont {Tokura}}]{Imada:1998}%
  \BibitemOpen
  \bibfield  {author} {\bibinfo {author} {\bibfnamefont {M.}~\bibnamefont
  {Imada}}, \bibinfo {author} {\bibfnamefont {A.}~\bibnamefont {Fujimori}}, \
  and\ \bibinfo {author} {\bibfnamefont {Y.}~\bibnamefont {Tokura}},\ }\href
  {\doibase 10.1103/RevModPhys.70.1039} {\bibfield  {journal} {\bibinfo
  {journal} {Rev. Mod. Phys.}\ }\textbf {\bibinfo {volume} {70}},\ \bibinfo
  {pages} {1039} (\bibinfo {year} {1998})}\BibitemShut {NoStop}%
\bibitem [{\citenamefont {Balents}(2010)}]{Balents:2010}%
  \BibitemOpen
  \bibfield  {author} {\bibinfo {author} {\bibfnamefont {L.}~\bibnamefont
  {Balents}},\ }\href {\doibase 10.1038/nature08917} {\bibfield  {journal}
  {\bibinfo  {journal} {Nature}\ }\textbf {\bibinfo {volume} {464}},\ \bibinfo
  {pages} {199} (\bibinfo {year} {2010})}\BibitemShut {NoStop}%
\bibitem [{\citenamefont {Cazalilla}\ \emph {et~al.}(2011)\citenamefont
  {Cazalilla}, \citenamefont {Citro}, \citenamefont {Giamarchi}, \citenamefont
  {Orignac},\ and\ \citenamefont {Rigol}}]{Cazalilla:2011}%
  \BibitemOpen
  \bibfield  {author} {\bibinfo {author} {\bibfnamefont {M.}~\bibnamefont
  {Cazalilla}}, \bibinfo {author} {\bibfnamefont {R.}~\bibnamefont {Citro}},
  \bibinfo {author} {\bibfnamefont {T.}~\bibnamefont {Giamarchi}}, \bibinfo
  {author} {\bibfnamefont {E.}~\bibnamefont {Orignac}}, \ and\ \bibinfo
  {author} {\bibfnamefont {M.}~\bibnamefont {Rigol}},\ }\href {\doibase
  10.1103/RevModPhys.83.1405} {\bibfield  {journal} {\bibinfo  {journal} {Rev.
  Mod. Phys.}\ }\textbf {\bibinfo {volume} {83}},\ \bibinfo {pages} {1405}
  (\bibinfo {year} {2011})}\BibitemShut {NoStop}%
\bibitem [{\citenamefont {Bloch}\ \emph {et~al.}(2008)\citenamefont {Bloch},
  \citenamefont {Dalibard},\ and\ \citenamefont {Zwerger}}]{Bloch:2008c}%
  \BibitemOpen
  \bibfield  {author} {\bibinfo {author} {\bibfnamefont {I.}~\bibnamefont
  {Bloch}}, \bibinfo {author} {\bibfnamefont {J.}~\bibnamefont {Dalibard}}, \
  and\ \bibinfo {author} {\bibfnamefont {W.}~\bibnamefont {Zwerger}},\ }\href
  {\doibase 10.1103/RevModPhys.80.885} {\bibfield  {journal} {\bibinfo
  {journal} {Rev. Mod. Phys.}\ }\textbf {\bibinfo {volume} {80}},\ \bibinfo
  {pages} {885} (\bibinfo {year} {2008})}\BibitemShut {NoStop}%
\bibitem [{\citenamefont {Trotzky}\ \emph {et~al.}(2008)\citenamefont
  {Trotzky}, \citenamefont {Cheinet}, \citenamefont {F\"{o}lling},
  \citenamefont {Feld}, \citenamefont {Schnorrberger}, \citenamefont {Rey},
  \citenamefont {Polkovnikov}, \citenamefont {Demler}, \citenamefont {Lukin},\
  and\ \citenamefont {Bloch}}]{Trotzky:2008a}%
  \BibitemOpen
  \bibfield  {author} {\bibinfo {author} {\bibfnamefont {S.}~\bibnamefont
  {Trotzky}}, \bibinfo {author} {\bibfnamefont {P.}~\bibnamefont {Cheinet}},
  \bibinfo {author} {\bibfnamefont {S.}~\bibnamefont {F\"{o}lling}}, \bibinfo
  {author} {\bibfnamefont {M.}~\bibnamefont {Feld}}, \bibinfo {author}
  {\bibfnamefont {U.}~\bibnamefont {Schnorrberger}}, \bibinfo {author}
  {\bibfnamefont {A.~M.}\ \bibnamefont {Rey}}, \bibinfo {author} {\bibfnamefont
  {A.}~\bibnamefont {Polkovnikov}}, \bibinfo {author} {\bibfnamefont {E.~A.}\
  \bibnamefont {Demler}}, \bibinfo {author} {\bibfnamefont {M.~D.}\
  \bibnamefont {Lukin}}, \ and\ \bibinfo {author} {\bibfnamefont
  {I.}~\bibnamefont {Bloch}},\ }\href {\doibase 10.1126/science.1150841}
  {\bibfield  {journal} {\bibinfo  {journal} {Science}\ }\textbf {\bibinfo
  {volume} {319}},\ \bibinfo {pages} {295} (\bibinfo {year}
  {2008})}\BibitemShut {NoStop}%
\bibitem [{\citenamefont {Nascimb\`{e}ne}\ \emph {et~al.}(2012)\citenamefont
  {Nascimb\`{e}ne}, \citenamefont {Chen}, \citenamefont {Atala}, \citenamefont
  {Aidelsburger}, \citenamefont {Trotzky}, \citenamefont {Paredes},\ and\
  \citenamefont {Bloch}}]{Nascimbene:2012}%
  \BibitemOpen
  \bibfield  {author} {\bibinfo {author} {\bibfnamefont {S.}~\bibnamefont
  {Nascimb\`{e}ne}}, \bibinfo {author} {\bibfnamefont {Y.-A.}\ \bibnamefont
  {Chen}}, \bibinfo {author} {\bibfnamefont {M.}~\bibnamefont {Atala}},
  \bibinfo {author} {\bibfnamefont {M.}~\bibnamefont {Aidelsburger}}, \bibinfo
  {author} {\bibfnamefont {S.}~\bibnamefont {Trotzky}}, \bibinfo {author}
  {\bibfnamefont {B.}~\bibnamefont {Paredes}}, \ and\ \bibinfo {author}
  {\bibfnamefont {I.}~\bibnamefont {Bloch}},\ }\href {\doibase
  10.1103/PhysRevLett.108.205301} {\bibfield  {journal} {\bibinfo  {journal}
  {Phys. Rev. Lett.}\ }\textbf {\bibinfo {volume} {108}},\ \bibinfo {pages}
  {205301} (\bibinfo {year} {2012})}\BibitemShut {NoStop}%
\bibitem [{\citenamefont {Bakr}\ \emph {et~al.}(2010)\citenamefont {Bakr},
  \citenamefont {Peng}, \citenamefont {Tai}, \citenamefont {Ma}, \citenamefont
  {Simon}, \citenamefont {Gillen}, \citenamefont {F\"{o}lling}, \citenamefont
  {Pollet},\ and\ \citenamefont {Greiner}}]{Bakr:2010}%
  \BibitemOpen
  \bibfield  {author} {\bibinfo {author} {\bibfnamefont {W.~S.}\ \bibnamefont
  {Bakr}}, \bibinfo {author} {\bibfnamefont {A.}~\bibnamefont {Peng}}, \bibinfo
  {author} {\bibfnamefont {M.~E.}\ \bibnamefont {Tai}}, \bibinfo {author}
  {\bibfnamefont {R.}~\bibnamefont {Ma}}, \bibinfo {author} {\bibfnamefont
  {J.}~\bibnamefont {Simon}}, \bibinfo {author} {\bibfnamefont {J.~I.}\
  \bibnamefont {Gillen}}, \bibinfo {author} {\bibfnamefont {S.}~\bibnamefont
  {F\"{o}lling}}, \bibinfo {author} {\bibfnamefont {L.}~\bibnamefont {Pollet}},
  \ and\ \bibinfo {author} {\bibfnamefont {M.}~\bibnamefont {Greiner}},\ }\href
  {\doibase 10.1126/science.1192368} {\bibfield  {journal} {\bibinfo  {journal}
  {Science}\ }\textbf {\bibinfo {volume} {329}},\ \bibinfo {pages} {547}
  (\bibinfo {year} {2010})}\BibitemShut {NoStop}%
\bibitem [{\citenamefont {Sherson}\ \emph {et~al.}(2010)\citenamefont
  {Sherson}, \citenamefont {Weitenberg}, \citenamefont {Endres}, \citenamefont
  {Cheneau}, \citenamefont {Bloch},\ and\ \citenamefont {Kuhr}}]{Sherson:2010}%
  \BibitemOpen
  \bibfield  {author} {\bibinfo {author} {\bibfnamefont {J.~F.}\ \bibnamefont
  {Sherson}}, \bibinfo {author} {\bibfnamefont {C.}~\bibnamefont {Weitenberg}},
  \bibinfo {author} {\bibfnamefont {M.}~\bibnamefont {Endres}}, \bibinfo
  {author} {\bibfnamefont {M.}~\bibnamefont {Cheneau}}, \bibinfo {author}
  {\bibfnamefont {I.}~\bibnamefont {Bloch}}, \ and\ \bibinfo {author}
  {\bibfnamefont {S.}~\bibnamefont {Kuhr}},\ }\href {\doibase
  10.1038/nature09378} {\bibfield  {journal} {\bibinfo  {journal} {Nature}\
  }\textbf {\bibinfo {volume} {467}},\ \bibinfo {pages} {68} (\bibinfo {year}
  {2010})}\BibitemShut {NoStop}%
\bibitem [{\citenamefont {Simon}\ \emph {et~al.}(2011)\citenamefont {Simon},
  \citenamefont {Bakr}, \citenamefont {Ma}, \citenamefont {Tai}, \citenamefont
  {Preiss},\ and\ \citenamefont {Greiner}}]{Simon:2011}%
  \BibitemOpen
  \bibfield  {author} {\bibinfo {author} {\bibfnamefont {J.}~\bibnamefont
  {Simon}}, \bibinfo {author} {\bibfnamefont {W.~S.}\ \bibnamefont {Bakr}},
  \bibinfo {author} {\bibfnamefont {R.}~\bibnamefont {Ma}}, \bibinfo {author}
  {\bibfnamefont {M.~E.}\ \bibnamefont {Tai}}, \bibinfo {author} {\bibfnamefont
  {P.~M.}\ \bibnamefont {Preiss}}, \ and\ \bibinfo {author} {\bibfnamefont
  {M.}~\bibnamefont {Greiner}},\ }\href {\doibase 10.1038/nature09994}
  {\bibfield  {journal} {\bibinfo  {journal} {Nature}\ }\textbf {\bibinfo
  {volume} {472}},\ \bibinfo {pages} {307} (\bibinfo {year}
  {2011})}\BibitemShut {NoStop}%
\bibitem [{\citenamefont {Kuklov}\ and\ \citenamefont
  {Svistunov}(2003)}]{Kuklov:2003}%
  \BibitemOpen
  \bibfield  {author} {\bibinfo {author} {\bibfnamefont {A.}~\bibnamefont
  {Kuklov}}\ and\ \bibinfo {author} {\bibfnamefont {B.}~\bibnamefont
  {Svistunov}},\ }\href {\doibase 10.1103/PhysRevLett.90.100401} {\bibfield
  {journal} {\bibinfo  {journal} {Phys. Rev. Lett.}\ }\textbf {\bibinfo
  {volume} {90}},\ \bibinfo {pages} {100401} (\bibinfo {year}
  {2003})}\BibitemShut {NoStop}%
\bibitem [{\citenamefont {Duan}\ \emph {et~al.}(2003)\citenamefont {Duan},
  \citenamefont {Demler},\ and\ \citenamefont {Lukin}}]{Duan:2003}%
  \BibitemOpen
  \bibfield  {author} {\bibinfo {author} {\bibfnamefont {L.-M.}\ \bibnamefont
  {Duan}}, \bibinfo {author} {\bibfnamefont {E.}~\bibnamefont {Demler}}, \ and\
  \bibinfo {author} {\bibfnamefont {M.}~\bibnamefont {Lukin}},\ }\href
  {\doibase 10.1103/PhysRevLett.91.090402} {\bibfield  {journal} {\bibinfo
  {journal} {Phys. Rev. Lett.}\ }\textbf {\bibinfo {volume} {91}},\ \bibinfo
  {pages} {090402} (\bibinfo {year} {2003})}\BibitemShut {NoStop}%
\bibitem [{\citenamefont {Garc\'{\i}a-Ripoll}\ and\ \citenamefont
  {Cirac}(2003)}]{GarciaRipoll:2003}%
  \BibitemOpen
  \bibfield  {author} {\bibinfo {author} {\bibfnamefont {J.~J.}\ \bibnamefont
  {Garc\'{\i}a-Ripoll}}\ and\ \bibinfo {author} {\bibfnamefont {J.~I.}\
  \bibnamefont {Cirac}},\ }\href {\doibase 10.1088/1367-2630/5/1/376}
  {\bibfield  {journal} {\bibinfo  {journal} {New J. Phys.}\ }\textbf {\bibinfo
  {volume} {5}},\ \bibinfo {pages} {76} (\bibinfo {year} {2003})}\BibitemShut
  {NoStop}%
\bibitem [{\citenamefont {Altman}\ \emph {et~al.}(2003)\citenamefont {Altman},
  \citenamefont {Hofstetter}, \citenamefont {Demler},\ and\ \citenamefont
  {Lukin}}]{Altman:2003a}%
  \BibitemOpen
  \bibfield  {author} {\bibinfo {author} {\bibfnamefont {E.}~\bibnamefont
  {Altman}}, \bibinfo {author} {\bibfnamefont {W.}~\bibnamefont {Hofstetter}},
  \bibinfo {author} {\bibfnamefont {E.}~\bibnamefont {Demler}}, \ and\ \bibinfo
  {author} {\bibfnamefont {M.~D.}\ \bibnamefont {Lukin}},\ }\href {\doibase
  10.1088/1367-2630/5/1/113} {\bibfield  {journal} {\bibinfo  {journal} {New J.
  Phys.}\ }\textbf {\bibinfo {volume} {5}},\ \bibinfo {pages} {113} (\bibinfo
  {year} {2003})}\BibitemShut {NoStop}%
\bibitem [{\citenamefont {Feynman}(1972)}]{Book:FeynmanStatMech1972}%
  \BibitemOpen
  \bibfield  {author} {\bibinfo {author} {\bibfnamefont {R.~P.}\ \bibnamefont
  {Feynman}},\ }\href@noop {} {\emph {\bibinfo {title} {{Statistical
  Mechanics}}}}\ (\bibinfo  {publisher} {W. A. Benjamin},\ \bibinfo {address}
  {Reading, MA},\ \bibinfo {year} {1972})\BibitemShut {NoStop}%
\bibitem [{\citenamefont {Endres}\ \emph {et~al.}(2011)\citenamefont {Endres},
  \citenamefont {Cheneau}, \citenamefont {Fukuhara}, \citenamefont
  {Weitenberg}, \citenamefont {Schauss}, \citenamefont {Gross}, \citenamefont
  {Mazza}, \citenamefont {Ba\~{n}uls}, \citenamefont {Pollet}, \citenamefont
  {Bloch},\ and\ \citenamefont {Kuhr}}]{Endres:2011}%
  \BibitemOpen
  \bibfield  {author} {\bibinfo {author} {\bibfnamefont {M.}~\bibnamefont
  {Endres}}, \bibinfo {author} {\bibfnamefont {M.}~\bibnamefont {Cheneau}},
  \bibinfo {author} {\bibfnamefont {T.}~\bibnamefont {Fukuhara}}, \bibinfo
  {author} {\bibfnamefont {C.}~\bibnamefont {Weitenberg}}, \bibinfo {author}
  {\bibfnamefont {P.}~\bibnamefont {Schauss}}, \bibinfo {author} {\bibfnamefont
  {C.}~\bibnamefont {Gross}}, \bibinfo {author} {\bibfnamefont
  {L.}~\bibnamefont {Mazza}}, \bibinfo {author} {\bibfnamefont {M.~C.}\
  \bibnamefont {Ba\~{n}uls}}, \bibinfo {author} {\bibfnamefont
  {L.}~\bibnamefont {Pollet}}, \bibinfo {author} {\bibfnamefont
  {I.}~\bibnamefont {Bloch}}, \ and\ \bibinfo {author} {\bibfnamefont
  {S.}~\bibnamefont {Kuhr}},\ }\href {\doibase 10.1126/science.1209284}
  {\bibfield  {journal} {\bibinfo  {journal} {Science}\ }\textbf {\bibinfo
  {volume} {334}},\ \bibinfo {pages} {200} (\bibinfo {year}
  {2011})}\BibitemShut {NoStop}%
\bibitem [{\citenamefont {Weitenberg}\ \emph {et~al.}(2011)\citenamefont
  {Weitenberg}, \citenamefont {Endres}, \citenamefont {Sherson}, \citenamefont
  {Cheneau}, \citenamefont {Schau\ss}, \citenamefont {Fukuhara}, \citenamefont
  {Bloch},\ and\ \citenamefont {Kuhr}}]{Weitenberg:2011}%
  \BibitemOpen
  \bibfield  {author} {\bibinfo {author} {\bibfnamefont {C.}~\bibnamefont
  {Weitenberg}}, \bibinfo {author} {\bibfnamefont {M.}~\bibnamefont {Endres}},
  \bibinfo {author} {\bibfnamefont {J.~F.}\ \bibnamefont {Sherson}}, \bibinfo
  {author} {\bibfnamefont {M.}~\bibnamefont {Cheneau}}, \bibinfo {author}
  {\bibfnamefont {P.}~\bibnamefont {Schau\ss}}, \bibinfo {author}
  {\bibfnamefont {T.}~\bibnamefont {Fukuhara}}, \bibinfo {author}
  {\bibfnamefont {I.}~\bibnamefont {Bloch}}, \ and\ \bibinfo {author}
  {\bibfnamefont {S.}~\bibnamefont {Kuhr}},\ }\href {\doibase
  10.1038/nature09827} {\bibfield  {journal} {\bibinfo  {journal} {Nature}\
  }\textbf {\bibinfo {volume} {471}},\ \bibinfo {pages} {319} (\bibinfo {year}
  {2011})}\BibitemShut {NoStop}%
\bibitem [{\citenamefont {Konno}(2005)}]{Konno:2005}%
  \BibitemOpen
  \bibfield  {author} {\bibinfo {author} {\bibfnamefont {N.}~\bibnamefont
  {Konno}},\ }\href {\doibase 10.1103/PhysRevE.72.026113} {\bibfield  {journal}
  {\bibinfo  {journal} {Phys. Rev. E}\ }\textbf {\bibinfo {volume} {72}},\
  \bibinfo {pages} {026113} (\bibinfo {year} {2005})}\BibitemShut {NoStop}%
\bibitem [{\citenamefont {Giamarchi}(2004)}]{Book:Giamarchi}%
  \BibitemOpen
  \bibfield  {author} {\bibinfo {author} {\bibfnamefont {T.}~\bibnamefont
  {Giamarchi}},\ }\href@noop {} {\emph {\bibinfo {title} {{Quantum Physics in
  one Dimension}}}}\ (\bibinfo  {publisher} {Oxford University Press},\
  \bibinfo {address} {Oxford, UK},\ \bibinfo {year} {2004})\BibitemShut
  {NoStop}%
\bibitem [{\citenamefont {Zvonarev}\ \emph {et~al.}(2007)\citenamefont
  {Zvonarev}, \citenamefont {Cheianov},\ and\ \citenamefont
  {Giamarchi}}]{Zvonarev:2007}%
  \BibitemOpen
  \bibfield  {author} {\bibinfo {author} {\bibfnamefont {M.~B.}\ \bibnamefont
  {Zvonarev}}, \bibinfo {author} {\bibfnamefont {V.~V.}\ \bibnamefont
  {Cheianov}}, \ and\ \bibinfo {author} {\bibfnamefont {T.}~\bibnamefont
  {Giamarchi}},\ }\href {\doibase 10.1103/PhysRevLett.99.240404} {\bibfield
  {journal} {\bibinfo  {journal} {Phys. Rev. Lett.}\ }\textbf {\bibinfo
  {volume} {99}},\ \bibinfo {pages} {240404} (\bibinfo {year}
  {2007})}\BibitemShut {NoStop}%
\bibitem [{\citenamefont {Schirotzek}\ \emph {et~al.}(2009)\citenamefont
  {Schirotzek}, \citenamefont {Wu}, \citenamefont {Sommer},\ and\ \citenamefont
  {Zwierlein}}]{Schirotzek:2009}%
  \BibitemOpen
  \bibfield  {author} {\bibinfo {author} {\bibfnamefont {A.}~\bibnamefont
  {Schirotzek}}, \bibinfo {author} {\bibfnamefont {C.-H.}\ \bibnamefont {Wu}},
  \bibinfo {author} {\bibfnamefont {A.}~\bibnamefont {Sommer}}, \ and\ \bibinfo
  {author} {\bibfnamefont {M.~W.}\ \bibnamefont {Zwierlein}},\ }\href {\doibase
  10.1103/PhysRevLett.102.230402} {\bibfield  {journal} {\bibinfo  {journal}
  {Phys. Rev. Lett.}\ }\textbf {\bibinfo {volume} {102}},\ \bibinfo {pages}
  {230402} (\bibinfo {year} {2009})}\BibitemShut {NoStop}%
\bibitem [{\citenamefont {Nascimb\`{e}ne}\ \emph {et~al.}(2009)\citenamefont
  {Nascimb\`{e}ne}, \citenamefont {Navon}, \citenamefont {Jiang}, \citenamefont
  {Tarruell}, \citenamefont {Teichmann}, \citenamefont {McKeever},
  \citenamefont {Chevy},\ and\ \citenamefont {Salomon}}]{Nascimbene:2009}%
  \BibitemOpen
  \bibfield  {author} {\bibinfo {author} {\bibfnamefont {S.}~\bibnamefont
  {Nascimb\`{e}ne}}, \bibinfo {author} {\bibfnamefont {N.}~\bibnamefont
  {Navon}}, \bibinfo {author} {\bibfnamefont {K.~J.}\ \bibnamefont {Jiang}},
  \bibinfo {author} {\bibfnamefont {L.}~\bibnamefont {Tarruell}}, \bibinfo
  {author} {\bibfnamefont {M.}~\bibnamefont {Teichmann}}, \bibinfo {author}
  {\bibfnamefont {J.}~\bibnamefont {McKeever}}, \bibinfo {author}
  {\bibfnamefont {F.}~\bibnamefont {Chevy}}, \ and\ \bibinfo {author}
  {\bibfnamefont {C.}~\bibnamefont {Salomon}},\ }\href {\doibase
  10.1103/PhysRevLett.103.170402} {\bibfield  {journal} {\bibinfo  {journal}
  {Phys. Rev. Lett.}\ }\textbf {\bibinfo {volume} {103}},\ \bibinfo {pages}
  {170402} (\bibinfo {year} {2009})}\BibitemShut {NoStop}%
\bibitem [{\citenamefont {Catani}\ \emph {et~al.}(2012)\citenamefont {Catani},
  \citenamefont {Lamporesi}, \citenamefont {Naik}, \citenamefont {Gring},
  \citenamefont {Inguscio}, \citenamefont {Minardi}, \citenamefont {Kantian},\
  and\ \citenamefont {Giamarchi}}]{Catani:2012}%
  \BibitemOpen
  \bibfield  {author} {\bibinfo {author} {\bibfnamefont {J.}~\bibnamefont
  {Catani}}, \bibinfo {author} {\bibfnamefont {G.}~\bibnamefont {Lamporesi}},
  \bibinfo {author} {\bibfnamefont {D.}~\bibnamefont {Naik}}, \bibinfo {author}
  {\bibfnamefont {M.}~\bibnamefont {Gring}}, \bibinfo {author} {\bibfnamefont
  {M.}~\bibnamefont {Inguscio}}, \bibinfo {author} {\bibfnamefont
  {F.}~\bibnamefont {Minardi}}, \bibinfo {author} {\bibfnamefont
  {A.}~\bibnamefont {Kantian}}, \ and\ \bibinfo {author} {\bibfnamefont
  {T.}~\bibnamefont {Giamarchi}},\ }\href {\doibase 10.1103/PhysRevA.85.023623}
  {\bibfield  {journal} {\bibinfo  {journal} {Phys. Rev. A}\ }\textbf {\bibinfo
  {volume} {85}},\ \bibinfo {pages} {023623} (\bibinfo {year}
  {2012})}\BibitemShut {NoStop}%
\bibitem [{\citenamefont {Koschorreck}\ \emph {et~al.}(2012)\citenamefont
  {Koschorreck}, \citenamefont {Pertot}, \citenamefont {Vogt}, \citenamefont
  {Fr\"{o}hlich}, \citenamefont {Feld},\ and\ \citenamefont
  {K\"{o}hl}}]{Koschorreck:2012}%
  \BibitemOpen
  \bibfield  {author} {\bibinfo {author} {\bibfnamefont {M.}~\bibnamefont
  {Koschorreck}}, \bibinfo {author} {\bibfnamefont {D.}~\bibnamefont {Pertot}},
  \bibinfo {author} {\bibfnamefont {E.}~\bibnamefont {Vogt}}, \bibinfo {author}
  {\bibfnamefont {B.}~\bibnamefont {Fr\"{o}hlich}}, \bibinfo {author}
  {\bibfnamefont {M.}~\bibnamefont {Feld}}, \ and\ \bibinfo {author}
  {\bibfnamefont {M.}~\bibnamefont {K\"{o}hl}},\ }\href {\doibase
  10.1038/nature11151} {\bibfield  {journal} {\bibinfo  {journal} {Nature}\
  }\textbf {\bibinfo {volume} {485}},\ \bibinfo {pages} {619} (\bibinfo {year}
  {2012})}\BibitemShut {NoStop}%
\bibitem [{\citenamefont {Palzer}\ \emph {et~al.}(2009)\citenamefont {Palzer},
  \citenamefont {Zipkes}, \citenamefont {Sias},\ and\ \citenamefont
  {K\"{o}hl}}]{Palzer:2009}%
  \BibitemOpen
  \bibfield  {author} {\bibinfo {author} {\bibfnamefont {S.}~\bibnamefont
  {Palzer}}, \bibinfo {author} {\bibfnamefont {C.}~\bibnamefont {Zipkes}},
  \bibinfo {author} {\bibfnamefont {C.}~\bibnamefont {Sias}}, \ and\ \bibinfo
  {author} {\bibfnamefont {M.}~\bibnamefont {K\"{o}hl}},\ }\href {\doibase
  10.1103/PhysRevLett.103.150601} {\bibfield  {journal} {\bibinfo  {journal}
  {Phys. Rev. Lett.}\ }\textbf {\bibinfo {volume} {103}},\ \bibinfo {pages}
  {150601} (\bibinfo {year} {2009})}\BibitemShut {NoStop}%
\bibitem [{\citenamefont {Johnson}\ \emph {et~al.}(2011)\citenamefont
  {Johnson}, \citenamefont {Clark}, \citenamefont {Bruderer},\ and\
  \citenamefont {Jaksch}}]{Johnson:2011}%
  \BibitemOpen
  \bibfield  {author} {\bibinfo {author} {\bibfnamefont {T.}~\bibnamefont
  {Johnson}}, \bibinfo {author} {\bibfnamefont {S.}~\bibnamefont {Clark}},
  \bibinfo {author} {\bibfnamefont {M.}~\bibnamefont {Bruderer}}, \ and\
  \bibinfo {author} {\bibfnamefont {D.}~\bibnamefont {Jaksch}},\ }\href
  {\doibase 10.1103/PhysRevA.84.023617} {\bibfield  {journal} {\bibinfo
  {journal} {Phys. Rev. A}\ }\textbf {\bibinfo {volume} {84}},\ \bibinfo
  {pages} {023617} (\bibinfo {year} {2011})}\BibitemShut {NoStop}%
\bibitem [{\citenamefont {Schecter}\ \emph {et~al.}(2012)\citenamefont
  {Schecter}, \citenamefont {Kamenev}, \citenamefont {Gangardt},\ and\
  \citenamefont {Lamacraft}}]{Schecter:2012}%
  \BibitemOpen
  \bibfield  {author} {\bibinfo {author} {\bibfnamefont {M.}~\bibnamefont
  {Schecter}}, \bibinfo {author} {\bibfnamefont {A.}~\bibnamefont {Kamenev}},
  \bibinfo {author} {\bibfnamefont {D.}~\bibnamefont {Gangardt}}, \ and\
  \bibinfo {author} {\bibfnamefont {A.}~\bibnamefont {Lamacraft}},\ }\href
  {\doibase 10.1103/PhysRevLett.108.207001} {\bibfield  {journal} {\bibinfo
  {journal} {Phys. Rev. Lett.}\ }\textbf {\bibinfo {volume} {108}},\ \bibinfo
  {pages} {207001} (\bibinfo {year} {2012})}\BibitemShut {NoStop}%
\bibitem [{\citenamefont {Spethmann}\ \emph {et~al.}()\citenamefont
  {Spethmann}, \citenamefont {Kindermann}, \citenamefont {John}, \citenamefont
  {Weber}, \citenamefont {Meschede},\ and\ \citenamefont
  {Widera}}]{Spethmann:2012}%
  \BibitemOpen
  \bibfield  {author} {\bibinfo {author} {\bibfnamefont {N.}~\bibnamefont
  {Spethmann}}, \bibinfo {author} {\bibfnamefont {F.}~\bibnamefont
  {Kindermann}}, \bibinfo {author} {\bibfnamefont {S.}~\bibnamefont {John}},
  \bibinfo {author} {\bibfnamefont {C.}~\bibnamefont {Weber}}, \bibinfo
  {author} {\bibfnamefont {D.}~\bibnamefont {Meschede}}, \ and\ \bibinfo
  {author} {\bibfnamefont {A.}~\bibnamefont {Widera}},\ }\href@noop {} {\
  }\Eprint {http://arxiv.org/abs/1204.6051v1} {arXiv:1204.6051v1} \BibitemShut
  {NoStop}%
\bibitem [{\citenamefont {Zvonarev}\ \emph {et~al.}(2009)\citenamefont
  {Zvonarev}, \citenamefont {Cheianov},\ and\ \citenamefont
  {Giamarchi}}]{Zvonarev:2009}%
  \BibitemOpen
  \bibfield  {author} {\bibinfo {author} {\bibfnamefont {M.}~\bibnamefont
  {Zvonarev}}, \bibinfo {author} {\bibfnamefont {V.}~\bibnamefont {Cheianov}},
  \ and\ \bibinfo {author} {\bibfnamefont {T.}~\bibnamefont {Giamarchi}},\
  }\href {\doibase 10.1103/PhysRevLett.103.110401} {\bibfield  {journal}
  {\bibinfo  {journal} {Phys. Rev. Lett.}\ }\textbf {\bibinfo {volume} {103}},\
  \bibinfo {pages} {110401} (\bibinfo {year} {2009})}\BibitemShut {NoStop}%
\bibitem [{\citenamefont {Ganahl}\ \emph {et~al.}(2012)\citenamefont {Ganahl},
  \citenamefont {Rabel}, \citenamefont {Essler},\ and\ \citenamefont
  {Evertz}}]{Ganahl:2012}%
  \BibitemOpen
  \bibfield  {author} {\bibinfo {author} {\bibfnamefont {M.}~\bibnamefont
  {Ganahl}}, \bibinfo {author} {\bibfnamefont {E.}~\bibnamefont {Rabel}},
  \bibinfo {author} {\bibfnamefont {F.}~\bibnamefont {Essler}}, \ and\ \bibinfo
  {author} {\bibfnamefont {H.}~\bibnamefont {Evertz}},\ }\href {\doibase
  10.1103/PhysRevLett.108.077206} {\bibfield  {journal} {\bibinfo  {journal}
  {Phys. Rev. Lett.}\ }\textbf {\bibinfo {volume} {108}},\ \bibinfo {pages}
  {077206} (\bibinfo {year} {2012})}\BibitemShut {NoStop}%
\bibitem [{\citenamefont {Gobert}\ \emph {et~al.}(2005)\citenamefont {Gobert},
  \citenamefont {Kollath}, \citenamefont {Schollw\"{o}ck},\ and\ \citenamefont
  {Sch\"{u}tz}}]{Gobert:2005}%
  \BibitemOpen
  \bibfield  {author} {\bibinfo {author} {\bibfnamefont {D.}~\bibnamefont
  {Gobert}}, \bibinfo {author} {\bibfnamefont {C.}~\bibnamefont {Kollath}},
  \bibinfo {author} {\bibfnamefont {U.}~\bibnamefont {Schollw\"{o}ck}}, \ and\
  \bibinfo {author} {\bibfnamefont {G.}~\bibnamefont {Sch\"{u}tz}},\ }\href
  {\doibase 10.1103/PhysRevE.71.036102} {\bibfield  {journal} {\bibinfo
  {journal} {Phys. Rev. E}\ }\textbf {\bibinfo {volume} {71}},\ \bibinfo
  {pages} {036102} (\bibinfo {year} {2005})}\BibitemShut {NoStop}%
\bibitem [{\citenamefont {Liang}\ \emph {et~al.}(2009)\citenamefont {Liang},
  \citenamefont {Kohn}, \citenamefont {Becker},\ and\ \citenamefont
  {Heinzen}}]{Liang:2009}%
  \BibitemOpen
  \bibfield  {author} {\bibinfo {author} {\bibfnamefont {J.}~\bibnamefont
  {Liang}}, \bibinfo {author} {\bibfnamefont {R.~N.}\ \bibnamefont {Kohn}},
  \bibinfo {author} {\bibfnamefont {M.~F.}\ \bibnamefont {Becker}}, \ and\
  \bibinfo {author} {\bibfnamefont {D.~J.}\ \bibnamefont {Heinzen}},\ }\href
  {\doibase 10.1364/AO.48.001955} {\bibfield  {journal} {\bibinfo  {journal}
  {Appl. Opt.}\ }\textbf {\bibinfo {volume} {48}},\ \bibinfo {pages} {1955}
  (\bibinfo {year} {2009})}\BibitemShut {NoStop}%
\bibitem [{\citenamefont {Vidal}(2004)}]{Vidal:2004}%
  \BibitemOpen
  \bibfield  {author} {\bibinfo {author} {\bibfnamefont {G.}~\bibnamefont
  {Vidal}},\ }\href {\doibase 10.1103/PhysRevLett.93.040502} {\bibfield
  {journal} {\bibinfo  {journal} {Phys. Rev. Lett.}\ }\textbf {\bibinfo
  {volume} {93}},\ \bibinfo {pages} {040502} (\bibinfo {year}
  {2004})}\BibitemShut {NoStop}%
\bibitem [{\citenamefont {Daley}\ \emph {et~al.}(2004)\citenamefont {Daley},
  \citenamefont {Kollath}, \citenamefont {Schollw\"{o}ck},\ and\ \citenamefont
  {Vidal}}]{Daley:2004}%
  \BibitemOpen
  \bibfield  {author} {\bibinfo {author} {\bibfnamefont {A.~J.}\ \bibnamefont
  {Daley}}, \bibinfo {author} {\bibfnamefont {C.}~\bibnamefont {Kollath}},
  \bibinfo {author} {\bibfnamefont {U.}~\bibnamefont {Schollw\"{o}ck}}, \ and\
  \bibinfo {author} {\bibfnamefont {G.}~\bibnamefont {Vidal}},\ }\href
  {\doibase 10.1088/1742-5468/2004/04/P04005} {\bibfield  {journal} {\bibinfo
  {journal} {J. Stat. Mech.: Theor. Exp.}\ ,\ \bibinfo {pages} {P04005}}
  (\bibinfo {year} {2004})}\BibitemShut {NoStop}%
\bibitem [{\citenamefont {White}\ and\ \citenamefont
  {Feiguin}(2004)}]{White:2004}%
  \BibitemOpen
  \bibfield  {author} {\bibinfo {author} {\bibfnamefont {S.~R.}\ \bibnamefont
  {White}}\ and\ \bibinfo {author} {\bibfnamefont {A.~E.}\ \bibnamefont
  {Feiguin}},\ }\href {\doibase 10.1103/PhysRevLett.93.076401} {\bibfield
  {journal} {\bibinfo  {journal} {Phys. Rev. Lett.}\ }\textbf {\bibinfo
  {volume} {93}},\ \bibinfo {pages} {76401} (\bibinfo {year}
  {2004})}\BibitemShut {NoStop}%
\bibitem [{\citenamefont {Verstraete}\ \emph {et~al.}(2004)\citenamefont
  {Verstraete}, \citenamefont {Garc\'{\i}a-Ripoll},\ and\ \citenamefont
  {Cirac}}]{Verstraete:2004}%
  \BibitemOpen
  \bibfield  {author} {\bibinfo {author} {\bibfnamefont {F.}~\bibnamefont
  {Verstraete}}, \bibinfo {author} {\bibfnamefont {J.~J.}\ \bibnamefont
  {Garc\'{\i}a-Ripoll}}, \ and\ \bibinfo {author} {\bibfnamefont {J.~I.}\
  \bibnamefont {Cirac}},\ }\href {\doibase 10.1103/PhysRevLett.93.207204}
  {\bibfield  {journal} {\bibinfo  {journal} {Phys. Rev. Lett.}\ }\textbf
  {\bibinfo {volume} {93}},\ \bibinfo {pages} {207204} (\bibinfo {year}
  {2004})}\BibitemShut {NoStop}%
\bibitem [{\citenamefont {Schollw\"{o}ck}(2005)}]{Schollwoeck:2005}%
  \BibitemOpen
  \bibfield  {author} {\bibinfo {author} {\bibfnamefont {U.}~\bibnamefont
  {Schollw\"{o}ck}},\ }\href {\doibase 10.1103/RevModPhys.77.259} {\bibfield
  {journal} {\bibinfo  {journal} {Rev. Mod. Phys.}\ }\textbf {\bibinfo {volume}
  {77}},\ \bibinfo {pages} {259} (\bibinfo {year} {2005})}\BibitemShut
  {NoStop}%
\end{thebibliography}%



\section*{Methods}

\subsection*{Simultaneous multiple-site addressing}  

We prepared the initial state with a single flipped spin at the centre of each one-dimensional (1D) system by using a new technique to address several lattice sites simultaneously. In our previous work, we used a single circular shaped Gaussian laser beam focussed onto a single lattice site together with a microwave field for addressing an individual spin in the lattice \mycite{Weitenberg:2011}. Instead of a circular beam, we now used arbitrary light-intensity patterns created with a spatial light modulator (SLM) of digital-micromirror-device (DMD) type in order to address several atoms at the same time. We generated a line-shaped light pattern by reflecting a Gaussian laser beam off the DMD (Texas Instruments, DLP Discovery 4100, 1024$\times$768 pixel, 13.7\,$\mu$m pixel size), and coupled it into the high-resolution imaging setup with a dichroic mirror. The magnification was such that one pixel of the DMD corresponds to $a_{\rm lat}/8 \approx 70$\,nm in the object plane. Compared to the diffraction limited spot size of $\sim600$\,nm this oversampling allowed us to implement an error-diffusion algorithm  \mycite{Liang:2009} to generate beam profiles with about 10\% root-mean-square variations of the peak intensity. Similar to our previous work \mycite{Weitenberg:2011}, we applied a feedback that shifted the pattern on the DMD for compensating slow phase drifts of the optical lattice.

The addressing laser had a wavelength 787.65\,nm and was $\sigma^-$ polarized in order to minimize the light shift for atoms in the initial state $\ket{F=1, m_F=-1}$, while creating a differential light shift between the initial state and $\ket{F=2, m_F=-2}$ of $\sim 35$\,kHz. In order to flip the spin, we used an HS1-pulse \mycite{Weitenberg:2011} of 20\,ms duration, $40$\,kHz sweep width and $-30$\,kHz frequency offset of the sweep centre from the bare resonance. The addressing beam was switched to full intensity for the spin-flip, and then back to 10\% of the intensity for pinning the spin impurity (see main text) with s-shaped ramps of 10\,ms duration. In relatively deep optical lattices ($V_x, V_y, V_z = 20\,E_r$) we obtained a spin-flip fidelity of $97(^{+2}_{-3})\%$, which is the same as in our previous experiment \mycite{Weitenberg:2011}. In this work, the lattice depths were kept at a lower value along the 1D chains ($V_x = 10\,E_r, V_y = 30\,E_r, V_z = 20\,E_r$) in order to reduce heating from the lattice ramp after the spin preparation, and we achieved a spin-flip fidelity of $88(5)\%$.

\subsection*{Numerical simulations}  In order to compute the time-dependent 1D many-body state
of the mobile impurity and the bath, we used numerical methods such as time-dependent matrix
product states (t-MPS, a.k.a.~time-dependent DMRG) \mycite{Vidal:2004,Daley:2004,White:2004,Verstraete:2004,Schollwoeck:2005} and simulated
the two-species Bose-Hubbard model of Eq.~(1) in the Supplementary Information.
For our parameters, considering MPS of dimension $200$ is sufficient to describe the impurity dynamics for $J/U \leq 0.25$,
while for $J/U>0.25$ a dimension of $400$ is more appropriate, as the subsystem entanglement grows significantly
close to and inside the SF  regime.
We incorporated the experimental conditions and preparation sequences as follows: The finite temperature of the initial MI state (experimentally
prepared with an effective $J/U=0.053$, at a lattice depth $V_x = 10\,E_r, V_y = 30\,E_r$) is considered
by randomly sampling product states of localized atoms from the grand-canonical
Gibbs-ensemble for the given initial $U$, with $J$ set to zero~\mycite{Sherson:2010}.

Deep in the MI regime, this is a very good approximation. Chemical potential $\mu$ and external confinement $\omega_{\rm trap}$ were determined from the
experiment~\mycite{Sherson:2010}. We generally used $100$ randomly drawn initial states, which are then subjected to time evolution in accordance with the parameter changes in the experiment. To obtain the thermal average of the impurity dynamics, the time-dependent local
densities  $\langle \hat n_{\downarrow,j}(t)\rangle$ for each initial state are used to form
an unweighted average.

When we simulated the system in the Heisenberg limit, deep in the MI regime, $J/U$ does not change during the state preparation and the time evolution. The atom on the central site is spin-flipped and then evolved in time together with the background.
When simulating the system at larger  $J/U$ than the initial one, we follow the experimental
sequence exactly. Correctly modeling the release of the spin impurity is essential, as the timescales for switching the addressing laser are similar to the tunneling times. Specifically, we
simulated the adiabatic state preparation as it is done in the experiment using
an initial ramp-down of the lattice depth $V_x$ within $50$\,ms according to
\begin{eqnarray}\label{eq:Vt}
   V_x(t)&&=V_i \nonumber \\
   &&+\frac{V_f-V_i}{2}\left[1+\coth\left(\frac{T}{\tau}\right)
            \tanh\left(\frac{2t-T}{\tau}\right) \right],
\end{eqnarray}
with $V_i=10\,E_r$,  $T=50$\,ms, $\tau=15$\,ms and $V_f$ the desired final lattice depth. We also incorporated into the time-dependent Hamiltonian parameters the
switching off of the addressing laser within 1\,ms.

\section*{Supplementary Information}

\renewcommand{\thefigure}{S\arabic{figure}}
\setcounter{figure}{0}

\subsection{The two-species Bose-Hubbard model and the Heisenberg model limit}

We described the effective dynamics in a 1D lattice by a two-species single-band Bose-Hubbard model \mycite{Cazalilla:2011}, which is valid at the temperatures, interaction strengths and lattice depths relevant to this study:
\begin{eqnarray}\label{eq:full2speciesHam}
  \hat  H_{\rm BH}
     &=&
    -J\sum_{\langle j, k\rangle,\sigma}
    \hat b^{\dagger}_{\sigma,j}
    \hat b_{\sigma,k}
    +\sum_{j,\sigma,\sigma'}\frac{U}{2}
    \hat n_{\sigma,j}(\hat n_{\sigma',j}-\delta_{\sigma,\sigma'}) \nonumber \\
     &&+\sum_{j,\sigma} (V_j^{\rm trap}-\mu) \hat n_{\sigma,j}.
\end{eqnarray}
Here, $\langle j,k \rangle, \sigma = \uparrow,\downarrow$ is the summation index comprising all neighboring sites $j$ and $k$ and  the two bosonic species \up\ and \down, $J$ is the single-particle tunneling rate, and $\hat b^{\dagger}_{\sigma,j}$ ($\hat b_{\sigma,j}$) is the creation (annihilation) operator for a boson of type $\sigma$ on lattice site $j$. The species-independent on-site interaction energy is denoted by $U$, whereas $\hat n_{\sigma,j}$ are the atomic number operators, $\delta_{\sigma,\sigma'}$ is the Kronecker symbol, and   $\mu$ is the chemical potential. The external harmonic confinement with trapping frequency $\omega_{\rm trap}/2\pi$ is described by $V_j^{\rm trap} = \tfrac{1}{2} m \omega_{\rm trap}^2 (j a_{\rm lat})^2$, where $m$ is the mass of $^{87}$Rb, and $a_{\rm lat}$ is the lattice constant.

In the limit of large interactions ($U\gg J$) and unity filling, it is possible to treat $\hat H_{\rm BH}$ in second order perturbation theory~\mycite{Kuklov:2003,Duan:2003,GarciaRipoll:2003,Altman:2003a}, mapping it to the isotropic spin-$1/2$ Heisenberg Hamiltonian [Eq.\,(1a)], by constructing effective spin-1/2 operators
\begin{subequations}
\begin{align}
  \hat S_j^x & =
    \frac{1}{2}\left( \hat b^{\dag}_{\uparrow,j} \hat b_{\downarrow,j}
                    +  \hat b^{\dag}_{\downarrow,j} \hat b_{\uparrow,j}
                \right) \\
  \hat S_j^y & =
    \frac{1}{2i}\left( \hat b^{\dag}_{\uparrow,j} \hat b_{\downarrow,j}
                    -  \hat b^{\dag}_{\downarrow,j} \hat b_{\uparrow,j}
                \right) \\
  \hat S_j^z & =
    \frac{1}{2}\left( \hat n_{\uparrow,j} -  \hat n_{\downarrow,j}
               \right)
\end{align}
\end{subequations}
and by introducing the exchange coupling $J_{\rm ex} = 4 J^2/U$.

In principle, the parabolic confinement $V_i$ makes superexchange site-dependent,
\begin{equation}\label{eq:}
    J_{{\rm ex}, j}=\frac{4J^2U}{U^2-(V_j-V_{j+1})^2}.
\end{equation}
However, given the relatively small system size (12 sites) and low trapping frequency $\omega_{\rm trap}/2\pi=85$\,Hz, the variation of $J_{\rm ex}$ across the system does not exceed $2\%$ at $V_x = 10\,E_r$. We can therefore safely consider the superexchange coupling constant, with $J_{{\rm ex},j} \approx J_{\rm ex}=4J^2/U$.

\subsection{Post-selecting samples}

\begin{figure}[!b]
    \centering
     \includegraphics[width=\columnwidth]{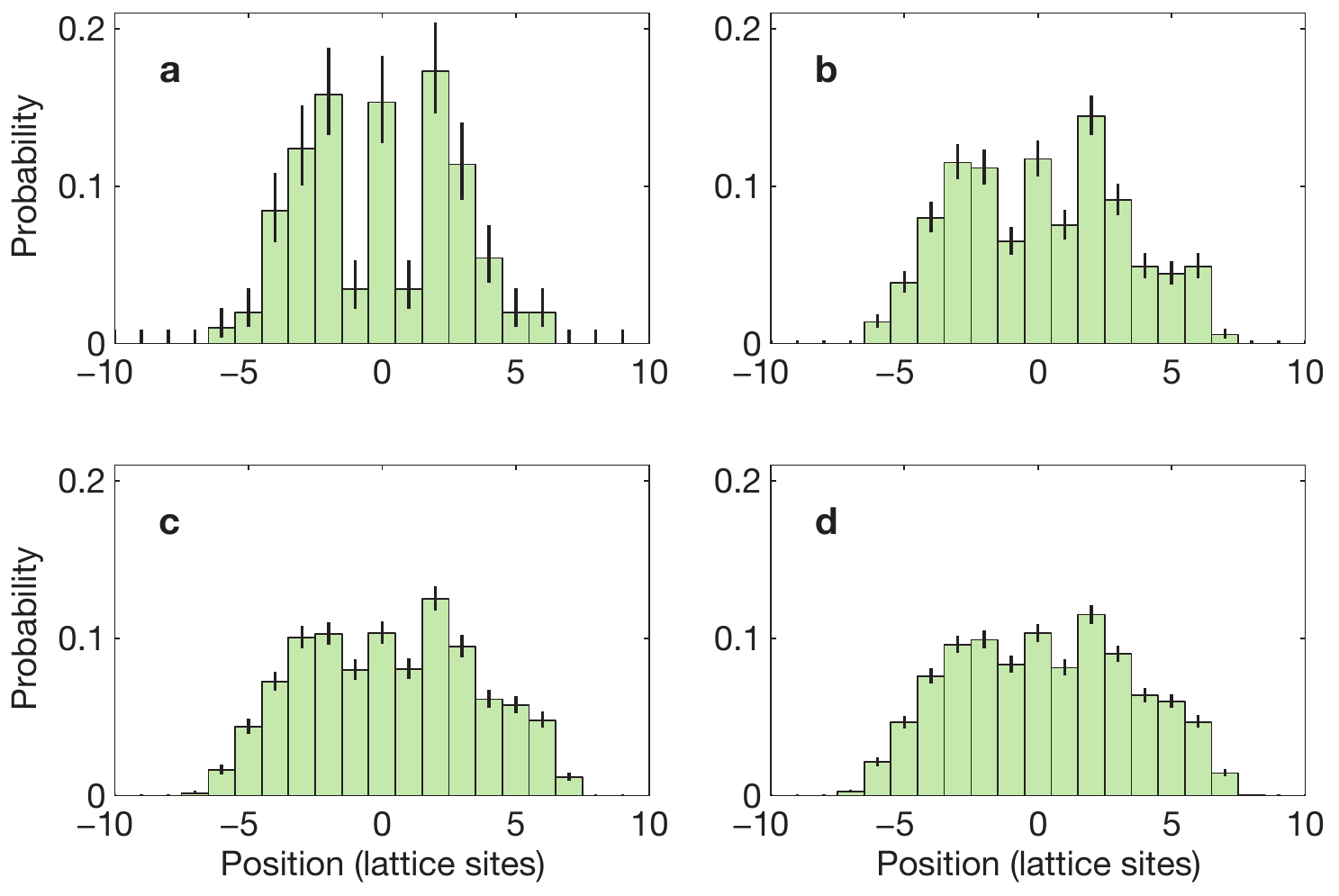}
    \caption{{\bf Post-selecting samples.} Shown are position distribution of the single-spin impurity for $J/U = 0.053$ and a hold time of $60$\,ms, after post-selecting samples with one ({\bf a}), less than three ({\bf b}), four ({\bf c}), and five ({\bf d}) empty sites in the chain.}
    \label{fig:post-selection}
  \end{figure}

In order to visualize the role of our post-selection method, we plotted the probability distributions of empty sites obtained from negative images including chains with more than one empty site (Fig.\,\ref{fig:post-selection}). We observe a clear decrease of the signal contrast as the maximum number of empty sites becomes larger. Although this is partially due to the temperature (which we discuss in the main text for positive images), the main reason for the decrease is that we cannot distinguish between spin impurities and excitations within the negative image.

The post-selection relies on the assumption that there is only one spin impurity in the system. This assumption is not valid in the  following two cases. First, the spin-flip on the initial site may fail due to the finite efficiency of the multi-site addressing scheme. In this case, if there is only one thermal excitation in the region of interest of the system, and we count this excitation as the spin impurity. Second, if there is a hole at the addressed site, no spin impurity is introduced into the system and the hole is mistakenly detected as the impurity. A hole is moving much faster than the impurity and after averaging over samples of different system size its quantum interference disappears over the relatively longer time scales on which spin-wave dynamics occurs. In both cases, we obtain distributions of thermal excitations, which have no clear interference-like structure.
What we obtained after post-selecting samples in our experiment is the sum of the position distribution of the spin impurity in the lower-temperature systems and the distribution of the thermal excitation. This leads to a decrease in contrast, which explains the small differences between the experimental data and the model in Fig.\,\ref{fig:spreading_speed}.

Furthermore, this mechanism for the reduction in contrast is why we claim in the manuscript
that the post-selected data should actually correspond to an even lower temperature (and thus higher contrast) than indicated in Fig.\,\ref{fig:excitation_effect}d. This remaining ambiguity of the spin-impurity position within the post-selected data could in principle be resolved by using direct spin-selective single-atom detection, which would allow for simultaneous imaging of both bath and impurity atoms. This, however, is experimentally challenging.

\subsection{Temperature effects}

\begin{figure}[!h]
    \centering
     \includegraphics[width=0.65\columnwidth]{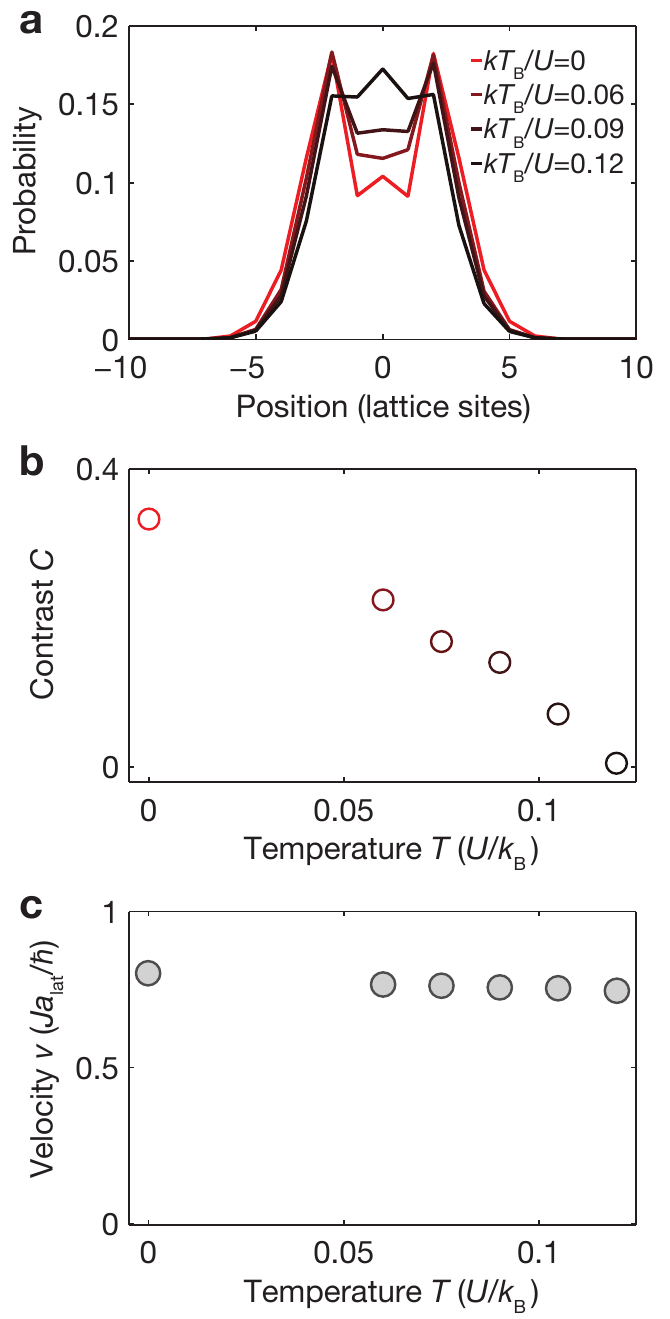}
    \caption{{\bf Effects of temperature on the dynamics.} ({\bf a}) Position distribution of the spin impurity for $J/U = 0.23$ and a hold time of $4$\,ms calculated with t-DMRG for different temperatures. Temperature dependence of the contrast ({\bf b}) and the spreading velocity ({\bf c}) extracted from the dataset of ({\bf a}).}
    \label{fig:temperature_effect}
  \end{figure}

We investigated the influence of temperature on the spin impurity dynamics for $J/U=0.23$ ($V_{x}=5\,E_r$), in addition to $J/U=0.053$ ($V_{x}=10\,E_r$) which was discussed in the main text. For this purpose, we numerically simulated the position distributions of the spin impurity after an evolution time of $4$\,ms for different temperatures (Fig.\,\ref{fig:temperature_effect}a). The simulation shows that even close to the transition point, the propagation velocity as a function of temperature remains constant whereas the contrast decreases (Fig.\,\ref{fig:temperature_effect}b and c), similar to the strongly interacting limit.

\end{document}